\documentclass[%
 reprint,
superscriptaddress,
 amsmath,amssymb,
 aps,
 floatfix,
 longbibliography,
]{revtex4-2}

\usepackage[dvipdfmx]{graphicx}
\usepackage{dcolumn}
\usepackage{bm}
\usepackage[mathlines]{lineno}
\usepackage{ulem}
\usepackage{xcolor}

\begin{document}

\preprint{APS/123-QED}

\title{Observation of spin splitting in the surface electronic structure of \\ antiferromagnet NdBi}

\author{Rikako Yamamoto}
 \email[]{ryama@hiroshima-u.ac.jp}
 \affiliation{%
 International Institute for Sustainability with Knotted Chiral Meta Matter (WPI‑SKCM$^{2}$), Hiroshima University, Higashi‑Hiroshima 739‑8526, Japan
}%
 \affiliation{%
Max Planck Institute for Chemical Physics of Solids, 01187 Dresden, Germany
}%
\author{Takeru Motoyama}
 \affiliation{%
 Graduate School of Advanced Science and Engineering, Hiroshima University, Higashi‑Hiroshima 739‑8526, Japan
}%

\author{Takuma Iwata}
 \affiliation{%
 International Institute for Sustainability with Knotted Chiral Meta Matter (WPI‑SKCM$^{2}$), Hiroshima University, Higashi‑Hiroshima 739‑8526, Japan
}%
 \affiliation{%
 Graduate School of Advanced Science and Engineering, Hiroshima University, Higashi‑Hiroshima 739‑8526, Japan
}%

\author{Towa Kosa}
\author{Yukimi Nishioka}
\author{Kazumasa Ideura}
 \affiliation{%
 Graduate School of Advanced Science and Engineering, Hiroshima University, Higashi‑Hiroshima 739‑8526, Japan
}%

\author{Masashi Arita}
 \affiliation{%
 Research Institute for Synchrotron Radiation Science (HiSOR), Hiroshima University, Higashi‑Hiroshima 739‑0046, Japann
}%

\author{Koji Miyamoto}
 \affiliation{%
 Research Institute for Synchrotron Radiation Science (HiSOR), Hiroshima University, Higashi‑Hiroshima 739‑0046, Japann
}%
\author{Taichi Okuda}
 \affiliation{%
 International Institute for Sustainability with Knotted Chiral Meta Matter (WPI‑SKCM$^{2}$), Hiroshima University, Higashi‑Hiroshima 739‑8526, Japan
}%
 \affiliation{%
 Research Institute for Synchrotron Radiation Science (HiSOR), Hiroshima University, Higashi‑Hiroshima 739‑0046, Japann
}%
\affiliation{%
 Research Institute for Semiconductor Engineering, Hiroshima University, Higashi-Hiroshima 739-8527, Japan
}%

\author{Akio Kimura}
 \affiliation{%
 International Institute for Sustainability with Knotted Chiral Meta Matter (WPI‑SKCM$^{2}$), Hiroshima University, Higashi‑Hiroshima 739‑8526, Japan
}%
 \affiliation{%
 Graduate School of Advanced Science and Engineering, Hiroshima University, Higashi‑Hiroshima 739‑8526, Japan
}%
\affiliation{%
 Research Institute for Semiconductor Engineering, Hiroshima University, Higashi-Hiroshima 739-8527, Japan
}%

\author{Takemi Yamada}
\author{Yuki Yanagi}
 \affiliation{%
 Liberal Arts and Sciences, Toyama Prefectural University, Imizu, 939-0398, Japan
}%

\author{Takahiro Onimaru}
 \affiliation{%
 Graduate School of Advanced Science and Engineering, Hiroshima University, Higashi‑Hiroshima 739‑8526, Japan
}%

\author{Kenta Kuroda}
 \affiliation{%
 International Institute for Sustainability with Knotted Chiral Meta Matter (WPI‑SKCM$^{2}$), Hiroshima University, Higashi‑Hiroshima 739‑8526, Japan
}%
 \affiliation{%
 Graduate School of Advanced Science and Engineering, Hiroshima University, Higashi‑Hiroshima 739‑8526, Japan
}%
\affiliation{%
 Research Institute for Semiconductor Engineering, Hiroshima University, Higashi-Hiroshima 739-8527, Japan
}%

\date{\today}

\begin{abstract}

Spin splitting in electronic band structures via antiferromagnetic orders is a new route to control spin-polarized carriers that is available for spintronics applications.
Here, we investigated the spin degree of freedom in the electronic band structures of the antiferromagnet NdBi using laser-based spin- and angle-resolved photoemission spectroscopy (laser-SARPES).
Our laser-SARPES experiments revealed that the two surface bands that appear in the antiferromagnetic state are spin-polarized in opposite directions as a counterpart of the spin splitting.
Moreover, we observed that the spin polarization is antisymmetric to the electron momentum, indicating that spin degeneracy is lifted due the breaking of inversion symmetry at the surface.
These results are well reproduced by our density functional theory calculations with the single-$q$ magnetic structure, implying that the spin-split surface state is determined by the breaking of inversion symmetry in concert with the antiferromagnetic order.

\end{abstract}

\maketitle



For several decades, the emergence of spin splitting in electronic band structures and the development of their control schemes have been continuing concerns because of its potential application, particularly in spintronics \cite{Manchon15,Soumyanarayanan16, He22}.
In general, two distinct pathways to generate the spin splitting are considered \cite{okuda13}.
One is the time-reversal ($\mathcal{T}$) symmetry breaking due to ferromagnetic order, leading to Zeeman effect.
In which case, the spin-polarization axis is aligned along the magnetization axis \cite{Petrovykh98}.
The other is the inversion ($\mathcal{P}$) symmetry breaking in combination with spin-orbit coupling (SOC).
It gives rise to Rashba-Dresselhaus effect, in which the spin splitting and the spin polarization axis depend sensitively on the electron momentum \cite{Dresselhaus55,Rashba59,Casella60, Bihlmayer22}.

In recent years, antiferromagnets have been attracting attention as a new platform for studying the spin-polarized electrons \cite{Jungwirth16,Baltz18}.
Even without net magnetization, in contrast to ferromagnets, antiferromagnetic (AFM) structures can induce spin splitting by either $\mathcal{T}$- or $\mathcal{P}$-symmetry breaking, which can be elucidated by the perspective from magnetic multipole \cite{Yatsushiro21,Kusunose22, Kusunose23,Hayami23} and altermagnetism \cite{Smejkal22,McClarty24}.
The new aspect of AFM states arising from the spin-split electronic structures have been intensively studied through the anomalous Hall effect \cite{Smejkal22_2}, spin-current generation \cite{Yang17,Hernandez21,Shao23} and anisotropic magnetoresistance \cite{Kriegner16}.
Moreover, the AFM order parameter (e.g., N\'eel vector) can act as a state variable for controlling these properties, which was recently demonstrated to be electrically switchable by the current-induced spin-orbit torque \cite{Wadley16,Nair20,Tsai20}.
However, research on antiferromagnets has just started,
and the difficulty in observing the electron spin splitting in the AFM states hinders the extent to which AFM order impacts electronic states. \cite{Zhu24,Krempasky24,Osumi24}.

The rare-earth monopnictides $RX$ ($R$: rare-earth, $X$: pnictogen) crystallizing in NaCl-type structure with the space group of $Fm\bar{3}m$ (No. 225, $O_{h}$) \cite{Duan07} have received renewed interest because their electronic structures can be manipulated by the AFM orders.
Indeed, the sensitivity of the electronic structures to the AFM structures \cite{Settai94, Kumigashira97,Takayama09,Oinuma19,Jang19,Kushnirenko22,Sakhya22,Li23} and the strong coupling of itinerant carriers with the 4$f$ electrons \cite{Kuroda20,Arai22} were observed by using angle-resolved photoemission spectroscopy (ARPES).
Particularly in NdBi, the emergence of a surface state and the evolution of the band splitting upon undergoing the AFM transition at $T_{\text{N}}$ = 24 K \cite{Tsuchida65} was reported \cite{Schrunk22,Honma23_2}.
Density functional theory (DFT) calculations suggested that it arises from the spin-split surface state formed inside the band-folding hybridization bulk gap \cite{Wang23}.
The spin information of the electronic structure has so far been deduced only by circular dichroism (CD) \cite{Schrunk22},
but, the CD signals in ARPES inherently detect the orbital degree of freedom rather than the spin information \cite{Park12,Ryu17,Sunko17}.
Therefore, utilizing spin-resolved ARPES (SARPES) is essential for reliably determining the spin splitting and its texture in mometnum space.

In this Letter, we used laser-based SARPES to directly investigate the spin degree of freedom in the electronic structure of the AFM state in NdBi.
Our experiments revealed that the band splitting of the surface states is due to the lifting of spin degeneracy.
In addition, since the observed spin polarizations were found to be antisymmetric with respect to the time-reversal invariant momentum, $\mathcal{P}$-breaking at the surface contributes to the spin-splitting.
These results were well reproduced by our DFT calculation, which also suggest that the AFM order contributes to the spin-polarized surface dispersion.

\begin{figure}
\includegraphics{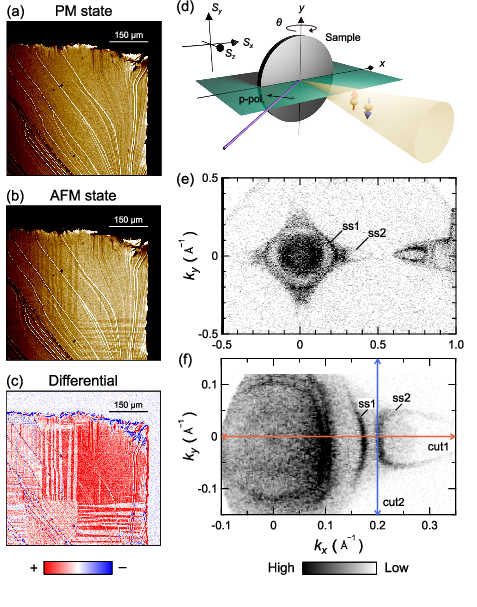}
\caption{\label{fig:fig1} (a), (b) Polarized microscope images of the cleaved (001) surface of NdBi in PM and AFM states. (c) The differential intensity image is obtained by subtracting the intensity between images of (a) and (b). (d) Schematic experimental geometry of our laser-SARPES and the definition of the resolved spin axis. (e) Fermi surface maps collected below $T_{\text{N}}$ using synchrotron-ARPES and (f) laser-ARPES.
}
\end{figure}

The single crystals of NdBi were synthesized by the Bi self-flux method as reported in literature \cite{Schrunk22}.
Synchrtron-based ARPES and laser-SARPES \cite{Iwata23} were performed in Research Institute for Synchrotron Radiation Science (HiSOR).
The energy resolution for synchrotron-based ARPES in BL9A was set to be below 20 meV.
For laser-SARPES, the photoelectrons excited by a micro-focused ultraviolet laser ($h\nu$ = 6.4 eV) with the $p$-polarization light were collected by a hemispherical electron analyzer (ScientaOmicron DA30L) and the spin polarization was determined by double very-low-energy-electron-diffraction (VLEED) detectors.
The experimental geometry and spin directions ($S_{x}$, $S_{y}$, $S_{z}$) are illustrated in Fig. \ref{fig:fig1}(d).
The energy and angular resolution of laser-ARPES (SARPES) were set to be 5 meV (25 meV) and 0.3$^{\circ}$ (0.75$^{\circ}$), respectively.
The samples were cooled to $\sim$7 K and cleaved in situ at a pressure less than $\sim5\times10^{-9}$ Pa to obtain the clean (001) surface.
Band calculations were performed using the DFT-based $ab\,initio$ calculation package WIEN2k \cite{Blaha20}, employing the PBE-GGA exchange-correlation potential with the DFT$+\,U$ method including SOC, where the onsite Hubbard-like Coulomb interactions were treated within the self-interaction correction \cite{Anisimov93} and the Coulomb parameter was set as $U=6.3$ eV and Hund's exchange interaction as $J=0.7$ eV only for 4$f$ electrons of each Nd atom.
The realistic tight-binding model for the AFM states of NdBi derived from the DFT $+ U$ calculation was constructed by Wannier90 \cite{Pizzi20}.
WannierTools \cite{Wu18} had been used to calculate the bulk and surface state spectra and study the impact of SOC and AFM order.

\begin{figure*}
\includegraphics{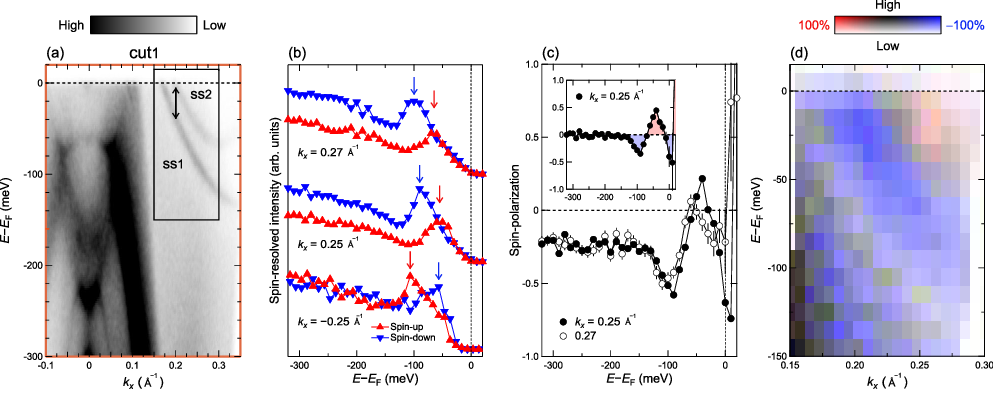}
\caption{\label{fig:fig2} (a) Band dispersion in the AFM state along $k_{x}$, corresponding to cut 1 in Fig. \ref{fig:fig1}(f). Double arrow represents energy gap between ss1 and ss2.
(b) Laser-SARPES spectra for the $S_y$ component at representative $k_{x}$ points, $-0.25$, 0.25 and 0.27 $\text{\AA}^{-1}$.
(c) Corresponding spin polarization spectra at $k_{x}$ = 0.25 and 0.27 $\text{\AA}^{-1}$.
The result after subtracting the spin-polarizartion background is presented in the inset.
(d) Spin-polarization map acquired in the $E-k_{x}$ window [rectangle denoted in (a)].
The blue-red color corresponds to the size of the spin polarization, and the brightness corresponds to the intensity \cite{Tusche15}.}
\end{figure*}

Most $RX$ compounds undergo an AFM transition, and their highly symmetric crystal structures allow for the existence of multiple magnetic domains \cite{Kuroda20, Arai22, Kushnirenko23, Honma23_2}. 
NdBi exhibits an AFM type-I structure \cite{Tsuchida65} and, therefore, multiple magnetic domains may exist as in isostructural compounds \cite{Honma23}. 
Figures \ref{fig:fig1}(a) and \ref{fig:fig1}(b) show polarized optical microscope images of the cleaved (001) surface in the paramagnetic (PM) and AFM states.
The image contrast changes across $T_{\text{N}}$, likely originating from the birefringence effect due to the tetragonal magnetic structure.
The spatial distribution of the domains becomes apparent in Fig. \ref{fig:fig1}(c), which was obtained by subtracting the images taken above and below $T_{\text{N}}$.
Notably, the domain size was large enough to resolve the AFM domains by our ARPES and SARPES, whose spatial resolutions are smaller than 50 $\mu$m.

Figure \ref{fig:fig1}(e) shows the Fermi surface in the AFM state measured by synchrotron-ARPES.
We observed the surface bands below $T_{\text{N}}$ along the $k_x$ axis but not along the $k_y$ \cite{Honma23_2}.
Here, we call the surface bands as ss1 and ss2.
The details of these Fermi surfaces are more clearly identified by our laser-ARPES in Fig. \ref{fig:fig1}(f).
The observed $C_2$ anisotropic electronic structure is consistent with the one reported previously, in which the magnetic moments are aligned in the in-plane \cite{Honma23_2}.

At the core of our motivation, we now move on to investigate the spin degrees of freedom of the bands appearing in the AFM state.
In this experiment, photoelectrons were detected along the high-symmetry line using $p$-polarized light [Fig. \ref{fig:fig1}(d)], ensuring the initial state information \cite{Kuroda16,Yaji17,Bentmann17}.
Figure \ref{fig:fig2}(a) displays the ARPES image along $k_{x}$ direction [cut 1 in Fig. \ref{fig:fig1}(f)].
The band dispersions were observed at $\sim$0.2 \AA$^{-1}$.
These two bands were previously considered counterparts of spin splitting because energy gap between ss1 and ss2 increases with decreasing temperature \cite{Schrunk22, Honma23_2}.
Thanks to the spin-resolution capability of our laser-SARPES, their spin information can now be accessed.
Figure \ref{fig:fig2}(b) shows the SARPES spectra at the representative $k_{x}$ points for $S_{y}$ corresponding to the helical spin component (see Fig. \ref{fig:fig1}(d)).
Examining the SARPES spectra taken at $k_{x}$ = 0.25 $\text{\AA}^{-1}$, it is recognized that the spin-up and spin-down spectra both show a single peak at different energies as indicated by arrows.
The same trend is also found at $k_{x}$ = 0.27 $\text{\AA}^{-1}$, but the peaks are shifted to lower energies, indicating that our laser-SARPES traces the band dispersion together with its spin-information.
As the two surface bands (ss1 and ss2) are assigned to the spin-up and spin-down states, we conclude that they originate from lifting of spin degeneracy.

Besides the spin splitting, negative spin polarization can be noticed in most of the energy range for both spin spectra taken at $k_{x} = 0.25$ and 0.27 \AA$^{-1}$ as displayed in Fig. \ref{fig:fig2}(c).
The inherent complexity of SARPES limits us to interpret the reason for this spin-polarized background, but it is probably due to spin-dependent photoemission matrix elements effect under $p$-polarized light excitation (see Fig.\ref{fig:fig1}(d)) \cite{Jozwiask11}.
Assuming that the background is independent of energy, we subtracted the offset component from the observed spin polarization to extract the initial state component.
The inset of Fig. \ref{fig:fig2}(c) shows the subtracted spin polarization spectra (see Fig. \ref{fig:sfig2} for more detail).
By applying the same analysis at each point, the spin polarization of ss1 and ss2 is displayed in spin-polarization band mapping as presented in Fig. \ref{fig:fig2}(d).

One may wonder about the origin of the spin splitting of ss1 and ss2 in the AFM state: Whether it is due to the \( \mathcal{T} \)-symmetry breaking or \( \mathcal{P} \)-symmetry breaking.
To verify these two possibilities, we collected laser-SARPES spectra at the opposite momentum, $k_{x} = -0.25$ $\text{\AA}^{-1}$ [bottom of Fig. \ref{fig:fig2}(b)].
Compared to the data for $k_{x} > 0$, the peak position in the spin-up and spin-down spectra at $k_{x} = -0.25$ $\text{\AA}^{-1}$ is swapped.
The spin state of ss1 and ss2 is, thus, found to be antisymmetric with respect to the invariant momentum ($k = 0$).
This characteristic feature indicates the consequence of the \( \mathcal{P} \)-symmetry breaking.
Let us note that the absence of the spin polarization offset for the negative $k_{x}$ in contrast to that for the positive $k_{x}$ may be due to the difference in experimental geometry: The incident angle was strongly deviated to take the laser-SARPES spectra for $\pm{k_x}$ [Fig. \ref{fig:fig1}(d)].

\begin{figure}
\includegraphics{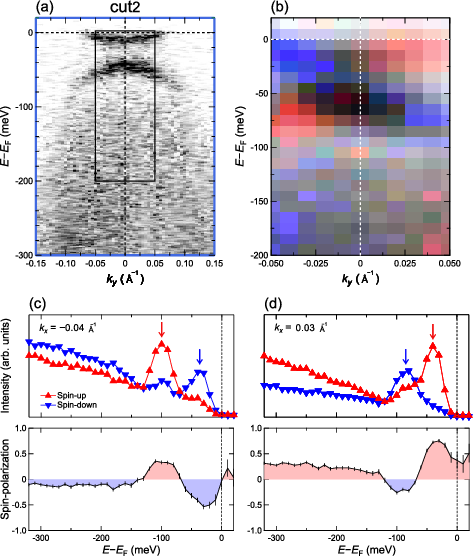}
\caption{\label{fig:fig3} (a) Laser-ARPES band map along $k_{y}$, corresponding to cut 2 in Fig. \ref{fig:fig1}(f).
(b) The spin-polarization map acquired in the area indicated by the solid rectangle shown in (a).
(c) and (d) Laser-SARPES spectra for the $S_x$ component at $k_{y}$=0.03 and $-$0.04 $\text{\AA}^{-1}$ and the corresponding spin polarization spectra.
}
\end{figure}
\begin{figure}[t]
\includegraphics[width=1.0\columnwidth]{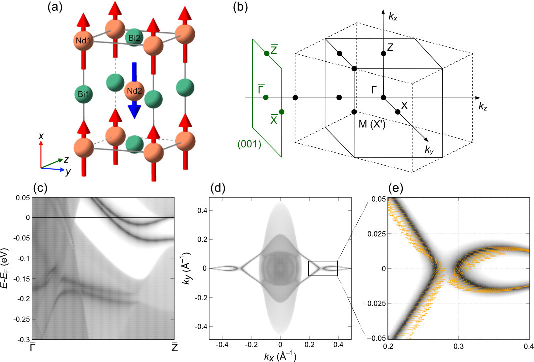}
\caption{\label{fig:fig4} (a) Single-$q$ AFM structure of NdBi, where the red and blue arrows indicate the directions of the Nd magnetic moments.
(b) Bulk Brillouin zone of single-$q$ and surface Brillouin zone of the (001) surface with the high-symmetry points. The tetragonal unit cell is shown by the dashed lines.
(c) Surface spectra along $\bar\Gamma$-$\bar{\text{Z}}$.
Here, we shifted the calculated $E_\text{F}$ to 40 meV higher in energy to agree with the experimental results. 
(d) Calculated Fermi surface of the surface states on the (001). 
(e) Magnified image of the Fermi surface with the spin texture indicated by yellow arrows.
}
\end{figure}

Our laser-SARPES can map out the spin texture to further understand the spin spitting.
Figure \ref{fig:fig3}(a) represents the band dispersion of ss1 and ss2 along $k_{y}$ direction [cut2 in Fig. \ref{fig:fig1}(e)].
The SARPES spectra of $S_{x}$ at $k_{y} = -0.04$ and 0.03 $\text{\AA}^{-1}$ and the corresponding spin polarization are presented in Figs. \ref{fig:fig3}(c) and \ref{fig:fig3}(d).
The spin polarization is reversed in ss1 and ss2, as counterparts of the spin splitting, and is antisymmetric with respect to $k_{y}$ = 0.
These trends of the spin-polarized bands are mapped out in Fig. \ref{fig:fig3}(b).
Notably, the magnitude of the spin polarization exceeds 50$\%$ even at small $k_y$.
This suggests that the spin polarization vector is aligned along $S_x$ direction, and thus, the spin texture largely deviates from the conventional helical one due to Rashba effect.

To gain insight into the spin-polarized surface states, we performed DFT calculations by considering the single-$q$ AFM structure.
We assume the tetragonal magnetic structure where the magnetic moment of Nd ions are aligned to the $x$ axis as shown in Fig. \ref{fig:fig4}(a).
Figure \ref{fig:fig4}(b) displays the three-dimensional bulk Brillouin zone of single-$q$ (dashed lines for the tetragonal unit cell) and the two-dimensional surface Brillouin zone of the (001) surface with the high-symmetry points.
The surface band dispersion along $k_{x}$, derived from the DFT calculation is displayed in Fig. \ref{fig:fig4}(c).
The corresponding Fermi surface of the (001) plane is displayed in Fig. \ref{fig:fig4}(d).
In this magnetic configuration, the sharp surface bands appear along $k_{x}$. 
The calculated spin texture in the rectangle area of Fig. \ref{fig:fig4}(d) is displayed in Fig. \ref{fig:fig4}(e).
The spin component is only in the $S_y$ direction at $k_y = 0$ while the spin component is mainly in the $S_x$ direction at $k_y \neq 0$ \cite{Wang23}.
This unconventional spin texture agrees with the laser-SARPES results presented in Figs. \ref{fig:fig2} and Figs. \ref{fig:fig3}.

In the series of $RX$, the reconstruction of the electronic bands in both the bulk and surface states occurs differently \cite{Kushnirenko22}.
We found that surface states observed in NdBi are strongly influenced not only by SOC but also by the AFM order of the 4$f$ electrons.
To discuss this point, we have constructed several tight-binding (TB) models with modified strength of SOC and Coulomb interaction $U,J$, and examined their effects on bulk and surface dispersions.
Here, we introduce two scaling parameters, $\alpha_{\rm SOC}$ and $\alpha_{U,J}$. The former generates a TB model with a SOC Hamiltonian multiplied by $\alpha_{\rm SOC}$, and the latter yields a TB model derived from a DFT+$U$ calculation with $(\alpha_{U,J}U,\alpha_{U,J}J)$.

\begin{figure*}[t!]
\includegraphics[scale=0.75]{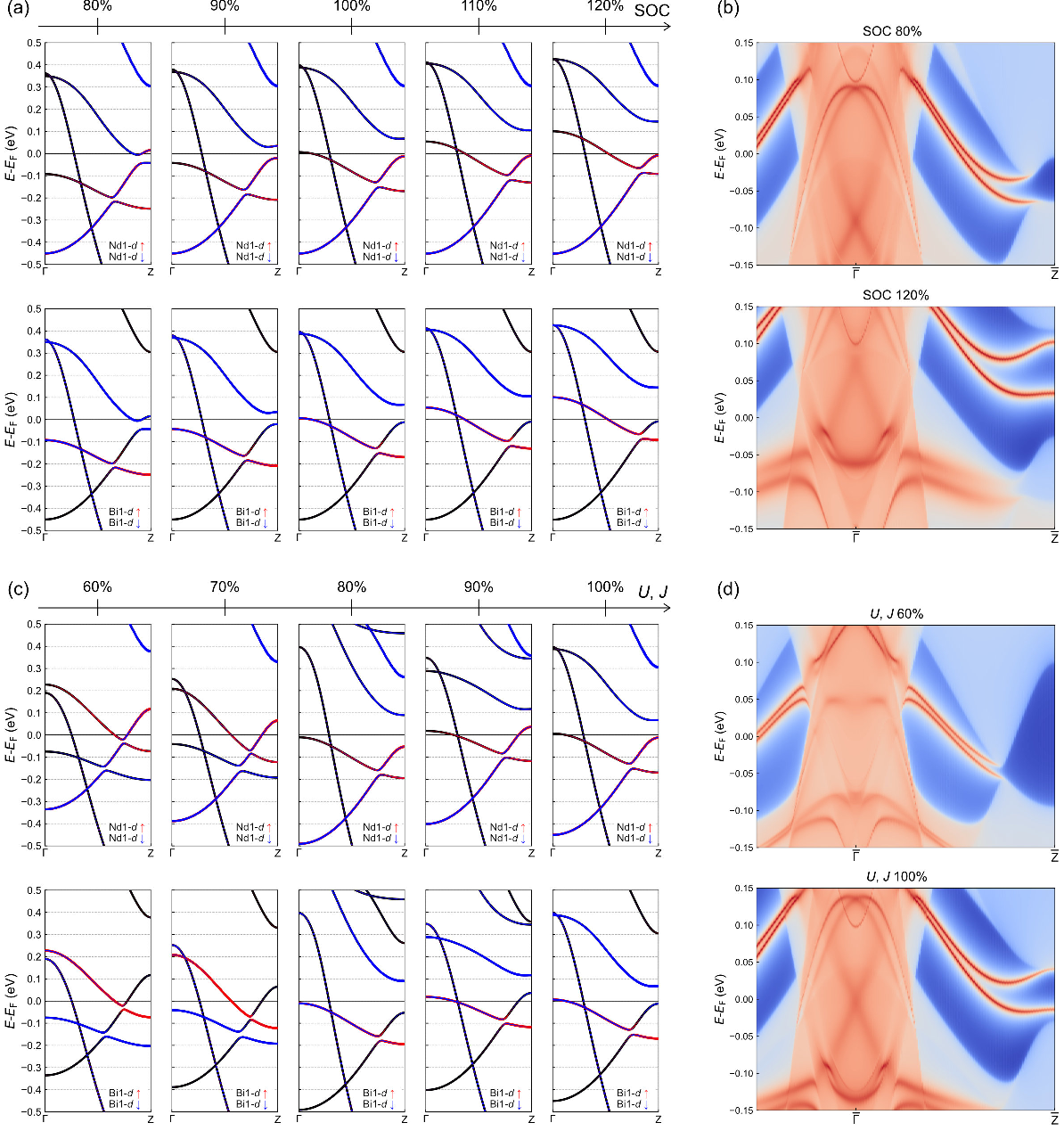}
\caption{\label{fig:fig5} Spin-polarization densities in the bulk bands of Nd1 (upper figures) and Bi1 (lower figures) in the AFM sublattice (see Fig. \ref{fig:fig4}(a)) for (a) the magnitude of SOC ($\alpha_{\rm SOC}$) and
(c) for the magnitude of $U$ and $J$ ($\alpha_{U,J}$).
Surface spectra along $\bar{\Gamma}$-$\bar{\text{Z}}$ with the bulk band projection for (b) $\alpha_{\rm SOC} = 80\%$ and $120\%$, and (d) $\alpha_{U,J} = 60\%$ and $100\%$, respectively.
}
\end{figure*}

Figure \ref{fig:fig5}(a) shows the bulk bands along the $\Gamma$-Z line for Nd1 and Bi1 (see Fig. \ref{fig:fig4}(a)) with varying $\alpha_{\rm SOC}$.
Here, $\alpha_{\rm SOC} = 100\%$ corresponds to the SOC value derived from our DFT calculations.
The data in Fig. \ref{fig:fig5}(a) clearly show a systematic evolution of the band gap near the Z point as $\alpha_{\rm SOC}$ increases. 
At $\alpha_{\rm SOC} = 120\%$, the conduction band bottom (CBB) and the valence band top (VBT) are located at $E-E_{\text{F}} \sim 0.15$ and 0 eV, respectively, forming a relatively large gap. 
In contrast, as decreasing $\alpha_{\rm SOC}$, the gap size decreases and at $\alpha_{\rm SOC} = 80\%$, the position of CBB and VBT is swapped, indicating band inversion.
These variations in the bulk band gap exert a impact on the surface band dispersions, as depicted in Fig. \ref{fig:fig5}(b).

The bulk bands with varying $U$ and $J$ values, scaled by $\alpha_{U,J}$, are displayed in Fig. \ref{fig:fig5}(c).
At $\alpha_{U,J} = 100\%$, the CBB at the Z point is located at $E-E_{\text{F}} \sim 0.1$ eV, lying higher in energy than the VBT at $E-E_{\text{F}} \sim 0$ eV.
However, at $\alpha_{U,J} = 60\%$, the CBB and VBT located at $E-E_{\text{F}} \sim -0.1$ eV and $+0.1$ eV, respectively, implying that the band inversion gap disappears at the Z point.
The impact of the bulk band inversion is clearly visible in the surface band dispersions presented in Fig. \ref{fig:fig5}(d). 
At $\alpha_{U,J} = 100\%$, the surface states appear over a wide $k$-region along the $\bar{\Gamma}$-$\bar{\text{Z}}$ line, while at $\alpha_{U,J} = 60\%$, the upper and lower surface states emerge from the bulk conduction and valence bands in between $\bar{\Gamma}$ and $\bar{\text{Z}}$ points. 
The splitting of the surface states emerges in both cases, but the energy gap and curvature of the surface varies, reflecting the variation in the size of the bulk projection gap.

The important finding of these calculations is that the surface bands are highly sensitive to the bulk bands near the Z point. 
The main reason for this is that surface bands are formed within the bulk projection gap, and the size and shape of this gap are determined by the bulk bands near the Z point.
Therefore, we conclude that the spin-split surface states are sensitively dependent not only on SOC but also on the AFM order through the coupling between the localized 4$f$ orbitals and the itinerant carriers, which provides a promising way to control spin-polarized carriers in antiferromagnets.

In summary, we investigated the spin and electronic structure of NdBi in the antiferromagnetic state by using laser-SARPES.
Our experiments have provided the conclusive evidence for spin splitting in the surface states.
Furthermore, by confirming the momentum dependence of the spin polarization, we showed that the spin-splitting originates from \(\mathcal{P}\)-symmetry breaking rather than \(\mathcal{T}\)-symmetry breaking.
The unconventional spin texture in the surface states were reproduced by the DFT theory considering the single-$q$ AFM state.
According to our DFT calculation, these surface states exist within the band gap determined by exchange splitting and would be sensitive to the magnetic order of the 4$f$ orbitals.

We thank Y. Fukushima for the discussions.
This work was also supported by JSPS KAKENHI (JP23K23211, JP23K17671, JP22H04483, JP22H01943, JP21H04652, and JP24K06943).

\appendix*
\section{\\Subtraction of the spin-polarization background}

The spin-resolved spectra and corresponding spin-polarization along $S_y$ in NdBi obtained by laser-SARPES are presented in Fig. \ref{fig:fig2} in the main manuscript.
There, not only the initial-state spin information of the surface states but also the spin-polarized signals coming from the final state effect are observed [58].
Here, we explain how both contributions are disentangled in our data.

In our laser-SARPES measurements, we used a very-low-energy-electron-diffracton (VLEED) type of spin detector [47].
The observed spin polarization of the photoelectron ($P$) is experimentally determined as follows:
\begin{equation} \label{eq:1}
P_{\rm{obs}}(E, \; k) = \frac{1}{S}\frac{I_{+}(E, \; k)-I_{-}(E, \; k)}{I_{+}(E, \; k)+I_{-}(E, \; k)},
\end{equation}
where $E$ is energy, $k$ is momentum, $I_{+}$ ($I_{-}$) is the photoelectron intensity for the positive (negative) magnetization of the VLEED target, and $S$ is the Sherman function of the detector (Here, we used $S =0.3$).
Eventually, the spin-resolved intensities, $I_{\uparrow}$ and $I_{\downarrow}$, can be obtained as follows:
\begin{equation} \label{eq:3}
\begin{split}
I_{\uparrow}(E, \; k) = \frac{1}{2}[1+P_{\rm{obs}}(E, \; k)]I_{\rm{tot}}(E, \; k)\\
I_{\downarrow}(E, \; k) = \frac{1}{2}[1-P_{\rm{obs}}(E, \; k)]I_{\rm{tot}}(E, \; k),
\end{split}
\end{equation}
where $I_{tot}$ is the total intensity corresponding to the sum of $I_{+}$ and $I_{-}$.

As shown in Fig. \ref{fig:fig2}(b) in the main manuscript, the spin splitting in the surface bands can be identified as the peak of each $I_{\uparrow}$ and $I_{\downarrow}$.
These spin states can also be discerned from the spin polarization spectra [Fig. \ref{fig:fig2}(c)].
However, it is observed that, in addition to this initial-state information, there is an additional spin polarization that makes the overall spin polarization negative.

\begin{figure}[t!]
\includegraphics{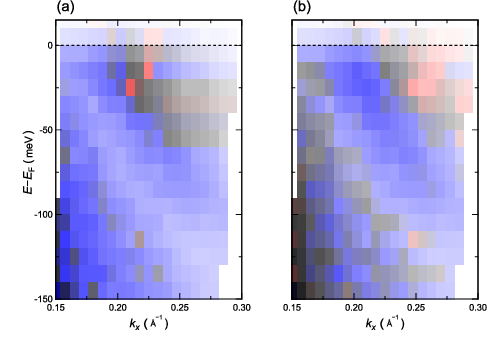}
\caption{\label{fig:sfig2} (a) Spin polarization ($P_{\rm{obs}}$) and (b) extracted initial state information ($P_{\rm{int}}$). The latter corresponds to Fig. 2(d) in the main text.
}
\end{figure}
Accordingly, the map of $P_{\rm{obs}}$ in Fig. \ref{fig:sfig2}(a) shows only the negative spin polarization in overall energy and momentum, which hinders the sign reversal feature of the initial state.

Assuming that the additional spin polarization coming from the final state is independent of energy and momentum, $P_{\rm{obs}}$ can be expressed as follows.
\begin{equation} \label{eq:4}
P_{\rm{obs}}(E, \; k) = P_{\rm{int}}(E, \; k)+P_{\rm{offset}},
\end{equation}
where $P_{\rm{int}}$ and $P_{\rm{offset}}$ is the spin polarization of the initial state and that of the final state effect.

Using $P_{\rm{offset}} =-0.23$, our analysis extracts $P_{\rm{int}}$, which well represents the sing-reversal feature of the spin polarization (see the inset of Fig.\ref{fig:fig2}(c)).
Eventually, $P_{\rm{int}}$ of the upper and lower bands are clearly seen in Figure \ref{fig:sfig2}(b) which is corresponding to Fig. \ref{fig:fig2}(d).\\

\providecommand{\noopsort}[1]{}\providecommand{\singleletter}[1]{#1}%

\end{document}